\documentclass[12pt]{article}

\normalbaselines
\begin{document}

\title{Accelerated expansion of the Universe \\ driven by dynamic self-interaction}
\author{A.B. Balakin\footnote{e-mail: Alexander.Balakin@ksu.ru}
\\ \it Department  of General Relativity and Gravitation,\\
\it Kazan State University, Kremlevskaya str. 18, Kazan 420008,
Russia \and H. Dehnen\footnote{e-mail: Heinz.Dehnen@uni-konstanz.de} \\ \it Universit\"at Konstanz, Fachbereich
Physik, \\ Fach M 677, D-78457, Konstanz, Germany }

\date{\today}

\maketitle

\begin{abstract}
We establish a new model, which takes into account a dynamic (inertial) self-interaction of gravitating
systems. The model is formulated by introduction  of a
new function depending on the square of the covariant derivative of the velocity four-vector of the
system as a whole into the Lagrangian. This term is meant for description of
both self-action of the system irregularly moving in the gravitational field, and
back-reaction of the motion irregularities on the gravity field.
We discuss one example of exact solution to the extended master equations in the framework
of cosmological model of the FLRW type with vanishing cosmological constant.
It is shown that accelerated expansion of the Universe
can be driven by traditional matter with positive pressure (e.g., dust, ultrarelativistic fluid)
due to the back-reaction of the gravity field induced by irregular motion of the system as a whole;
this back-reaction is shown to be characterized by the negative effective pressure.

\end{abstract}

\noindent
PACS number(s): 04.40.-b ; 98.80.Jk ; 04.20.Jb \\

\noindent
Keyword(s): irregular motion, accelerated expansion,
alternative to Dark Energy.

\newpage

\section{Introduction}\noindent

The observational fact that the Universe expands with acceleration is till now a
puzzle for theoretical cosmology. One can formulate three main ideas, on which the explanations of this
phenomenon could be based. The first idea is that there exists some exotic
substratum, Dark Energy, possessing a negative effective pressure (see, e.g., \cite{DE1,DE2,DE3}
for review and references).
The second one is focused on modifications of the so-called geometric sector of relativistic theory of gravity:
one can mention, for instance, $f(R)$ theory, the Gauss-Bonnet model, etc. (see, e.g.,
\cite{fR1,fR2,fR3} for review and references). The Lagrangians of such theories contain the invariants constructed
with nonlinear combinations of the Ricci scalar and the Riemann and Ricci tensors. The third
idea introduces interactions of new types between gravity on the one hand, and
fields and matter on the other hand (see, e.g., \cite{NM1,NM2,NM3,Obukhov}). The Non-minimal Field Theory is
the most elaborated trend in this direction, and the corresponding models
can be described by introduction of cross-invariants into the Lagrangian, which contain all admissible
convolutions of the Riemann, Ricci tensors and Ricci scalar in combinations with the field strength tensors
(see, e.g., \cite{NM2,NM3,BDZ}).

The theory of dynamic self-interaction
of gravitating systems, which we establish here, is in line with the third idea.
The motivation of the dynamic extention of the gravitational
theory has two aspects: mathematical and physical ones. Started from the mathematical
point of view, one can see that pure geometrical objects, which we use in the
Lagrangian of the gravity field (the Riemann, Ricci tensors and Ricci
scalar), contain second order partial derivatives of the metric. Since the covariant derivative of
the metric itself is considered to be equal to zero, there are no geometric invariants
containing the first derivatives of the metric only. Moreover, the covariant
derivative of the scalar field reduces to the partial derivative, the field
strength tensors for the electromagnetic and gauge fields, being the
skew-symmetric quantities, in fact do not contain the Christoffel symbols, the four-divergence of the
vector potential is assumed to be vanishing due to the Lorentz gauge.
Thus, the Lagrangians of the Einstein-Maxwell, Einstein-Yang-Mills-Higgs, etc. theories
do not contain invariants, into which only the first derivatives of the metric enter. The
situation changes essentially, when it is acceptable to introduce into the Lagrangian
the invariants containing the covariant derivative of the velocity
four-vector, attributed to the macroscopic motion of the system as a whole.
In this case one deals with the non-vanishing first derivatives of the metric, and
such a gravitational theory can be regarded as a modification of the Einstein theory of a new type.
The mathematical arguments supporting such a modification can be supplemented by the physical
ones.
Indeed, an accelerated point-like electrically  charged particle produces
electromagnetic radiation and is influenced by the back-reaction force,
proportional to the second derivative of its velocity \cite{LL}. This is a
clear example of dynamic self-interaction fulfilled by means of the electromagnetic
field, the particle acceleration being the necessary condition of such
self-interaction. It is well-known also, that irregularities of a medium motion lead to
the dynamo-optical phenomena, which can be described by including
covariant derivatives of the velocity four-vector into the Lagrangian of the moving
electrodynamic systems in combination with the Maxwell tensor \cite{LL}.
The back-reaction of irregularly moving electrodynamic system on the gravity
field was discussed in \cite{AlpBal}.
A question arises: is it possible to describe
a self-acceleration / deceleration of a material system as a result of a
gravitational back-reaction on its irregular motion? We show below that it is indeed
possible, when the Lagrangian of a matter contains some complementary invariant constructed
as a square of the covariant derivative of the velocity four-vector of a
system as a whole. As a cosmological consequence of this theory  we obtain
exact solution, which describes accelerated expansion of the Universe filled
with traditional matter with positive pressure in the case, when the lambda-term is vanishing.

\section{Dynamic extention of the gravitational theory}

Let us consider the variation of the action functional
\begin{equation}
S[g_{ik}] = \int d^4 x \sqrt{-g}\
\left[\frac{R}{2\kappa} + L_{({\rm m})} + F(\Psi^2)
\right]\,, \label{act}
\end{equation}
with respect to metric $g_{ik}$. The determinant of the metric, $g = {\rm det}(g_{ik})$,
the Ricci scalar, $R$, and the Lagrangian of a matter, $L_{({\rm m})}$, are
the standard elements of the action in the Einstein theory of gravity \cite{MTW}.
The new element in (\ref{act}) is the term $F(\Psi^2)$, which is presented
by a generic function of the argument $\Psi^2$
\begin{equation}
\Psi^2 \equiv \Psi_{ik} \Psi^{ik} \,, \quad \Psi_{ik} \equiv
\nabla_i U_k  \,, \label{act2}
\end{equation}
introducing a covariant derivative $\nabla_i U_k$ of the time-like
four-vector field $U^k$. We assume that $U^k$ is normalized by unity, i.e., $g_{ik}U^iU^k=1$, and
is interpreted as a velocity four-vector of the matter. More precisely, we
consider $U^k$ as a time-like eigen-vector of the stress-energy tensor of
the matter $T^{({\rm m})}_{ik}$ (the definition given by Landau and Lifshitz)
\begin{equation}
T^{({\rm m})}_{ik} U^k  = W U_i   \,, \quad T^{({\rm m})}_{ik} \equiv - \frac{2}{\sqrt{-g}} \frac{\delta
\left[\sqrt{-g} \ L_{({\rm m})}\right]}{\delta g^{ik}}\,.
 \label{eigen}
\end{equation}
The eigen-value $W$ relates to the scalar of the energy density of the matter,
and the stress-energy tensor of a matter $T^{({\rm m})}_{ik}$
can be now decomposed as
\begin{equation}
T^{({\rm m})}_{ik} = W U_i U_k +  {\cal
P}_{ik} \,, \quad {\cal P}_{ik} = - P \Delta_{ik} + \Pi_{ik} \,. \label{T11}
\end{equation}
Here ${\cal P}_{ik}$ is an anisotropic pressure tensor
containing a sum of the Pascal (isotropic) pressure, $P$, and a non-equilibrium pressure tensor, $\Pi_{ik}$.
The tensor $\Delta_{ik} \equiv g_{ik} {-} U_iU_k$ is a projector.
Thus, the first (algebraic) relation in (\ref{eigen}) allows us
to use one scalar $W$, one four-vector $U^k$ and one tensor ${\cal P}_{ik}$
instead of ten components $T^{({\rm m})}_{ik}$. The alternative approach
based on the Eckart definition of the macroscopic velocity four-vector
involves an additional particle number four-vector $N^k$, a heat flux
four-vector $q^k$ and the corresponding thermodynamic relations, thus
sophisticating the model.

The tensor $\Psi_{ik}$ can also be decomposed into the sum of
irreducible parts: acceleration four-vector $DU^{i}$,
shear tensor $\sigma_{ik}$, rotation tensor $\omega_{ik}$, and
expansion scalar $\Theta$
\begin{equation}
\Psi_{ik} = \nabla_i U_k = U_i DU_k + \sigma_{ik} + \omega_{ik} +
\frac{1}{3} \Delta_{ik} \Theta \,. \label{act3}
\end{equation}
These basic quantities are defined as
$$
DU_k \equiv U^m \nabla_m U_k \,,
\quad  \sigma_{ik} \equiv \frac{1}{2}\Delta_i^m \Delta_k^n
\left(\nabla_m U_n + \nabla_n U_m \right) - \frac{1}{3}\Delta_{ik}
\Theta \,,
$$
\begin{equation}
\omega_{ik} \equiv \frac{1}{2}\Delta_i^m \Delta_k^n \left(\nabla_m
U_n - \nabla_n U_m \right) \,, \quad \Theta \equiv \nabla_m U^m
\,. \label{act4}
\end{equation}
The quantity $\Psi^2$, the argument of the function $F(\Psi^2)$, has the form
\begin{equation}
\Psi^2 = DU_k DU^k + \sigma_{ik} \sigma^{ik} + \omega_{ik}
\omega^{ik} + \frac{1}{3}\Theta^2 \,. \label{act5}
\end{equation}
The variation of the action functional (\ref{act}) with respect to
the metric yields
\begin{equation}
R_{ik} - \frac{1}{2} R \ g_{ik} =  \kappa \left(T^{({\rm
m})}_{ik} + {\cal T}_{ik} \right) \,. \label{Ein}
\end{equation}
The tensor $T^{({\rm
m})}_{ik}$ is defined by (\ref{eigen}) and (\ref{T11}). The
new term ${\cal T}_{ik}$, defined as
\begin{equation}
{\cal T}_{ik} \equiv - \frac{2}{\sqrt{-g}} \frac{\delta
\left[\sqrt{-g} \ F(\Psi^2) \right]}{\delta g^{ik}}\,,
\label{TmD}
\end{equation}
can be represented in the following form
$$
{\cal T}_{ik} =
g_{ik} F(\Psi^2)  + 2\frac{d^2F}{d(\Psi^2)^2} \left[ U_{(i} \Psi_{k)m}  -  \Psi_{(ik)} U_m \right] \nabla^m \Psi^2 +
$$
\begin{equation}
+ 2\frac{dF}{d\Psi^2} \left[ \left(\Psi_{[mi]} \Psi_k^{\ m} +  \Psi_{[mk]}
\Psi_i^{\ m} \right)+
U_{(i} \nabla_{k)} \Theta + U_{(i} R_{k)m}U^m - (D + \Theta)\Psi_{(ik)} \right]
\,.
\label{Tpsi}
\end{equation}
As usual, the symbols
$\Psi_{(im)}$ and $\Psi_{[im]}$
denote symmetrization and skew-symmetrization, respectively.
Since the velocity four-vector is normalized, i.e., $g_{ik}U^iU^k
= 1$, the variation procedure with respect to metric $g_{ik}$ is
supplemented by the variation of the velocity:
\begin{equation}
\delta U^i = \frac{1}{4} \delta g^{pq}\left( U_p \delta^i_q + U_q
\delta^i_p \right)\,, \label{aux1}
\end{equation}
(see, e.g., \cite{CQG07} for details).
The sum of two tensors $T^{({\rm m})}_{ik}$ and
${\cal T}_{ik}$ should be divergence-free due to the Bianchi
identities
\begin{equation}
\nabla^k \left[T^{({\rm m})}_{ik} + {\cal T}_{ik} \right] =
0 \,. \label{0sKK17}
\end{equation}
To make the model complete, one should
formulate constitutive equations for $P$ and $\Pi_{ik}$. Here we restrict
ourselves by the ansatz, that the Pascal pressure is given by a barotropic function
$P=P(W)$, and the non-equilibrium pressure is a function of the energy density,
acceleration four-vector, shear tensor and expansion
scalar, i.e., $\Pi_{ik}{=}\Pi_{ik}(W, DU^l, \sigma_{mn}, \Theta)$.
Thus, we obtain a self-consistent model.

\section{Application to the space-time of the FLRW type}

\subsection{Key equations}

Let us assume the metric to have the following form
\begin{equation}
ds^2 = dt^2 - a^2(t)\left(dx^2 + dy^2 + dz^2 \right) \,,
\label{metric}
\end{equation}
and the velocity four-vector to have only one component $U^k =
\delta^k_0$. We deal with the so-called synchronous frame of
references. The tensor $\Psi_{ik}$ for this case
\begin{equation}
\Psi_{ik} = \nabla_i U_k = - \Gamma^0_{ik} = \frac{1}{2} \dot{g}_{ik} = \frac{\dot{a}}{a}\left(g_{ik}-U_i U_k\right) \,,
\label{F1}
\end{equation}
is symmetric. The dot denotes the ordinary derivative with respect
to time. The decomposition (\ref{act3}) has now only one
non-vanishing irreducible component
\begin{equation}
\Psi_{ik} = \frac{1}{3} \Delta_{ik} \Theta = \frac{\dot{a}}{a} \Delta_{ik} = H(t) \Delta_{ik} \,, \label{F2}
\end{equation}
and the scalar $\Psi^2$ reduces to $3H^2$. The stress-energy tensor (\ref{Tpsi})
converts into
\begin{equation}
{\cal T}^i_{k}[H] = \delta^i_k \left[ f(H) - H \frac{df}{dH}
\right] + \frac{1}{3} \left(\delta^i_0 \delta^0_k - \delta^i_k
\right) \ \dot{H} \frac{d^2f}{dH^2} \,, \label{Ttheta2}
\end{equation}
where a new unknown function $f(H) \equiv F(3H^2)$ is introduced.
Surprisingly, the four-divergence of the tensor (\ref{Ttheta2})
vanishes for arbitrary $f(H)$, i.e., $\nabla^k{\cal T}_{ik}[H] =
0$. Thus, we obtain from (\ref{0sKK17}) that $\nabla^k T^{({\rm m})}_{ik}= 0$ also,
and the energy and momentum of the matter conserve separately.
Clearly, the velocity four-vector $U^i=\delta^i_0$ is the time-like
eigenvector of this tensor, related to the eigen-value
\begin{equation}
\omega(H) = f(H) - H \frac{df}{dH} \,. \label{wH}
\end{equation}
Consequently, for the model under consideration the velocity four-vector $U^i$
happens to be an eigen-vector for the total stress-energy tensor $T^{({\rm m})}_{ik} + {\cal
T}_{ik}$, the corresponding eigen-value being the sum $W+\omega(H)$. Other three
eigen-values coincide and are equal to
\begin{equation}
\pi_{(1)} = \pi_{(2)} = \pi_{(3)} \equiv \pi(H) = - \omega(H)
+ \frac{1}{3}  \dot{H} \frac{d^2f}{dH^2}
\,.  \label{pH}
\end{equation}
One can interpret the quantities $\omega(H)$ and $\pi(H)$ as an effective (inertial) energy density and
effective (inertial) pressure, respectively, which are produced by irregularities of the
macroscopic motion of the system as a whole. The signs of these quantities
can be positive or negative depending on the choice of the function $f(H)$.
Let us note that if $f(H)$ is linear, say,
$f(H) = 3 A H$, then $H \frac{df}{dH} = f(H)$, $\frac{d^2f}{dH^2}=0$, and this function
disappears from the key equation, since $\omega(H)=\pi(H)=0$. Clearly, for such a case the
corresponding terms in the action functional
\begin{equation}
\int d^4x \sqrt{-g} A \Theta = \int d^4x A
\partial_k \left(\sqrt{-g} U^k \right) \,, \label{Ein7}
\end{equation}
gives a complete divergence and disappears at the variation
procedure.

The stress-energy tensor of matter (\ref{T11}) inherits the
symmetry of the FLRW space-time, thus, ${\cal P}_{ik} = - {\cal P}(t) \Delta_{ik}$, where ${\cal P}(t)$ can generally
include both Pascal ($P$) and  non-equilibrium ($\Pi$) parts, ${\cal P}(t)=P(t)+\Pi(t)$.
One can reduce the equations for the gravity field to the
following two equations
\begin{equation}
3H^2 =  \kappa \left[ W + f(H) - H \frac{df}{dH} \right]
\,, \label{Ein1}
\end{equation}
\begin{equation}
\dot{H}  = - \frac{1}{2} \kappa \left(  {\cal P} + W + \frac{1}{3} \dot{H} \frac{d^2f}{dH^2} \right) \,.
\label{Ein2}
\end{equation}
Using a differentiation of the first Einstein equation
(\ref{Ein1}) and excluding $\dot{H}$ from (\ref{Ein2}), we can
obtain the standard balance equation for the energy and momentum
of the matter
\begin{equation}
\dot{W} + 3 H (W+{\cal P})  = 0 \,, \label{balance}
\end{equation}
which, indeed, does not contain information about the function
$f(H)$.

Let us consider an exactly solvable model in which the lambda-term is considered to
be vanishing and the equation of state is
presented by the standard barotropic formula
\begin{equation}
{\cal P}  = (\gamma -1) W \,, \label{EOS}
\end{equation}
thus assuming that the Pascal (equilibrium) pressure, $P$, is proportional
to the energy density $W$, and the non-equilibrium pressure $\Pi$ is absent.
As usual, for such an equation of state one obtains from (\ref{balance})
\begin{equation}
\frac{P(t)}{(\gamma-1)} = W(t) = W_0 \left[\frac{a(t)}{a(t_0)}
\right]^{-3\gamma} \,. \label{W}
\end{equation}
In order to obtain $H(t)$ we have to solve the key equation
\begin{equation}
3H^2  + \kappa \left[H f^{\prime}(H) - f(H) \right] =
\kappa W_0 \left[\frac{a(t)}{a(t_0)} \right]^{-3\gamma} \,.
\label{key0}
\end{equation}
The acceleration parameter $-q(t)$
\begin{equation}
-q(t) \equiv \frac{\ddot{a}}{aH^2} = 1 + \frac{\dot{H}}{H^2}
\label{key15}
\end{equation}
can be presented in terms of $f(H)$ and its derivatives as
\begin{equation}
-q[f(H(t))] = 1 - \gamma \ \frac{\left\{3H^2 + \kappa [H f^{\prime}(H) -
f(H)] \right\}}{ 2H^2 \left[1 + \frac{\kappa}{6} f^{\prime \prime}(H) \right]} \,.
\label{key16}
\end{equation}
This formula is a direct consequence of (\ref{Ein2}), (\ref{Ein1}) and
(\ref{EOS}). When $f(H)=0$, the acceleration parameter $-q = 1-\frac{3\gamma}{2}$ is negative
for the traditional matter with $1 \leq \gamma \leq \frac{4}{3}$, as it should be, nevertheless,
the situation changes significantly, when  $f(H) \neq 0$.
The form of the function $f(H)$ is a subject of modeling. In the next papers we
hope to consider this function in
context of fitting of observational data, but here we restrict ourselves by
the analysis of one model only, which allows us to find an example of exact
solution and to study the acceleration parameter explicitly.

\subsection{Example of exact solution}

We consider the function $F(\Psi^2)$ to be quadratic function of its
argument, i.e., the function $f(H)=F(3H^2)$ has the form
\begin{equation}
\kappa f(H) = \alpha + \beta H^2 + \sigma H^4 \,.
\label{Ein8}
\end{equation}
According to (\ref{Ein1}) the constant $\alpha$ redefines the cosmological
constant, and we assume that $\alpha=0$.
The second parameter, $\beta$, can
be associated with the initial value of the Hubble function $H(t_0)=H_{0}$:
if we put $\kappa f(H)$ from (\ref{Ein8})
into (\ref{key0}), we obtain at $t=t_0$, that $\beta {=} {-} 3(1{+}\sigma
H^2_{0}) {+} \frac{\kappa W_0}{H^2_0}$.
Then (\ref{key0}) transforms into the bi-quadratic equation
\begin{equation}
H^4 - H^2 \left(H^2_0 - \frac{\kappa W_0}{3\sigma H^2_0} \right)
- \frac{\kappa W_0}{3\sigma} \left[\frac{a(t)}{a(t_0)}\right]^{-3\gamma} =
0\,,
\label{Ein9}
\end{equation}
which gives the following solutions
\begin{equation}
H^2(t) = \frac{1}{2}\left(H^2_0 - \frac{\kappa W_0}{3\sigma H^2_0} \right)
\pm
\sqrt{ \frac{1}{4}\left(H^2_0 - \frac{\kappa W_0}{3\sigma H^2_0} \right)^2 +
\frac{\kappa W_0}{3\sigma} \left[\frac{a(t)}{a(t_0)}\right]^{-3\gamma}} \,.
\label{Ein10}
\end{equation}
Now we have to specify the signs of the parameters $\beta$ and $\sigma$. We
assume here that $\sigma$ is positive and $\beta$ is negative, thus, the
function $\kappa f(H)$ has one local maximum at $H_{({\rm max})}=0$ and two
minimums at $H_{({\rm min})} = \pm \sqrt{-\frac{\beta}{2\sigma}}$. We assume
also, that the initial energy density $W_0$ is restricted by inequality
$\kappa W_0 \leq 3\sigma H^4_0$. Then the sign minus in front of the square root
corresponds to imaginary solution for $H(t)$. At $a \to \infty$ the second (real) solution
yields the asymptotic value of the Hubble function
\begin{equation}
H^2_{\infty} = H_0^2 - \frac{\kappa W_0}{3\sigma H_0^2}  \,, \quad H^2_0 \geq \frac{1}{2}H^2_{\infty} \,.
\label{Ein19}
\end{equation}
When $H_0 = \left( \frac{\kappa W_0}{3\sigma}\right)^{\frac{1}{4}}$, this
parameter vanishes, i.e., $H_{\infty}=0$.

Now the parameter of acceleration can be presented in the form
\begin{equation}
-q(t) = \frac{(4-3\gamma) H^2(t) + (3\gamma -2)H^2_{\infty}}{ 2\left[2H^2(t) -
H^2_{\infty}\right]}\,.
\label{Ein14}
\end{equation}
Clearly, the quantity $-q(t)$ is non-negative for the traditional matter with $1 \leq \gamma \leq
\frac{4}{3}$, i.e., we deal with the accelerated expansion. When $H_{\infty} \neq
0$, the acceleration parameter tends to one ($-q(\infty) = 1$) in the asymptotic limit $t \to
\infty$, and the scale factor is of the de Sitter form $a(t) \propto \exp[H_{\infty}t]$.
When $H_{\infty} = 0$, the solution for the Hubble function is
\begin{equation}
H = H_0 \left[\frac{a(t)}{a(t_0)}\right]^{- \frac{3\gamma}{4}}  \,,
\label{Ein91}
\end{equation}
the scale factor $a(t)$ takes the form
\begin{equation}
a(t) = a(t_0)  \left[1+ \frac{3}{4}\gamma H_0 (t-t_0)\right]^{\frac{4}{3\gamma}}
\,,
\label{Ein92}
\end{equation}
and the acceleration parameter becomes constant
\begin{equation}
-q(t) = 1-\frac{3}{4} \gamma \,,
\label{Ein93}
\end{equation}
being, of course, non-negative for the traditional matter with $1 \leq \gamma \leq
\frac{4}{3}$.

The effective (inertial) energy density $\omega(t)$ (\ref{wH}) takes now the form
\begin{equation}
\omega(t) = 3H^2(t) - \kappa W_0 \left[\frac{a(t)}{a(t_0)}\right]^{-3\gamma}
\,,
\label{Ein151}
\end{equation}
and tends asymptotically to the non-negative quantity $\omega(\infty) = 3H^2_{\infty}$.
The effective (inertial) pressure $\pi(t)$ (\ref{pH}) satisfies the relation
\begin{equation}
\pi(t) + \omega(t) =  - \frac{1}{3} H^2(t) [1+q(t)] f^{\prime \prime}(H)
\,.
\label{Ein152}
\end{equation}
In the asymptotic regime $-q \to 1$, thus, $\pi(\infty) = - \omega(\infty) =
- 3H^2_{\infty}$ is negative. In other words, the dynamic self-interaction influences the gravitating
system analogously to the effective (depending on time) lambda-term, and this explicit analogy can explain
the fact of accelerated expansion of the Universe. When
the Universe reaches the de Sitter stage, this analogy
becomes complete, since the constant value $3H^2_{\infty}$ can be considered
as effective (asymptotic) cosmological constant $\Lambda_{\infty}$.

\section{Conclusions}

\noindent
1. We established a new model, which accounts for dynamic (inertial) self-interaction of gravitating
systems.  The extension of the theory of gravity is based on the introduction of some unknown function $F(\Psi^2)$
depending on square of covariant derivative $\Psi_{ik} = \nabla_i U_k$ of the velocity
four-vector $U^k$ of the system as a whole. This unknown function should be determined using astrophysical
and cosmological observations, as well as gravitational experiments. Since
$F(\Psi^2)$ is the function of velocity derivatives, the Newtonian limit of
this modified theory of gravity is not violated, i.e., all the supplementary
terms in the Lagrangian decomposition start with $(v/c)^n$, $n \geq1$ and vanish at $c \to
\infty$.

\noindent
2. In this Letter we discuss only one example of exact solution to the extended master equations
attributed to the isotropic cosmological model with vanishing cosmological constant.
In the case, when the function $F(\Psi^2)$ is quadratic in its argument, we have shown explicitly that
the accelerated expansion of the Universe can be driven by matter with
positive pressure (e.g., dust, ultrarelativistic fluid), when the dynamic (inertial) self-interaction of
the gravitating system is taken into account.

\noindent
3. The dynamic self-interaction of the gravitating systems results in the
appearance of additional effective (inertial) contributions to the total energy density $\omega$ and
pressure $\pi$ (see (\ref{wH}), (\ref{pH}), (\ref{Ein151}) and
(\ref{Ein152})). These contributions look like the consequences of an effective lambda-term, which
generally depends on time, but becomes constant on the de Sitter stage of
the Universe evolution. In the asymptotic regime $\omega$ becomes positive and $\pi$ negative, so that their
sum vanishes.  Thus, we can consider the accelerated expansion of the
Universe to be driven by dynamic (inertial) self-interaction, which is
characterized by negative pressure induced by back-reaction of the gravity
field on the irregular motion of the system as a whole.

\noindent
4. In the context of isotropic cosmology only one type of the irregularity of motion, namely,
expansion ($\nabla_k U^k \neq 0$), contributes into the dynamic
self-interaction. For the static spherically symmetric gravitating systems the non-vanishing acceleration
four-vector $DU_k$ should be taken into account. For the systems with
pp-wave symmetry the shear tensor $\sigma_{ik}$ plays the main role in such
dynamic self-interaction. As for rotating systems, such as spiral
galaxies and pulsars, the contribution of the rotation tensor $\omega_{ik}$ seems to be the main one.

\noindent
5. The theory of dynamic self-interaction of an irregularly moving
gravitating system should be experimentally tested. We could suggest at least three
ways, how to do it. First, we intend to construct a realistic
model of the Universe accelerated expansion by fitting the function $-q[f(H(t))]$ (see (\ref{key16})), based on
the cosmological data; the reconstruction of the function $f(H)$ will give
us the basic function $F(\Psi^2)$. Second, we hope to analyse post-Newtonian
effects in this theory and to consider the corresponding experimental tests in the Solar
system. Third, one can consider the flat rotation velocity curves of the
spiral galaxies in the context of theory of dynamic self-interaction.
We hope to study the mentioned models in the nearest future.

\section*{Acknowledgments}\noindent

\noindent
The work was supported by the Deutsche Forschungsgemeinschaft
(Grant No. 436RUS113/487/0-5), and partially by the
Russian Foundation for Basic Research (Grants No.
08-02-00325-a and 09-05-99015).

\end{document}